\documentclass[11pt,twoside]{article}
\usepackage{asp2010}

\resetcounters

\begin{document}

\title{The Far-IR View of Star and Planet Forming Regions}
\author{Javier R. Goicoechea$^1$, M. Audard$^2$, C. Joblin$^3$, I. Kamp$^4$, M.R. Meyer$^5$
\affil{$^1$Centro de Astrobiolog\'{\i}a (CSIC-INTA), Madrid, Spain.}
\affil{$^2$Department of Astronomy, University of Geneva, Switzerland.}
\affil{$^3$Universit\'e de Toulouse, UPS-OMP-CNRS, IRAP, Toulouse, France.}
\affil{$^4$Kapteyn Astronomical Institute, Groningen, The Netherlands.}
\affil{$^5$Institute of Astronomy, ETH Zurich, Switzerland.}}

\begin{abstract}

The far-IR range is a  critical wavelength range to characterize the physical 
and chemical processes that transform the interstellar material into stars and planets. 
Objects in the earliest phases of stellar and planet evolution  release most of their energy 
at these long wavelengths. 
In this contribution we briefly summarise some of the most relevant scientific advances achieved
by the  \textit{Herschel Space Observatory} in the field. We also anticipate  those  that will be made possible
by the large increase in sensitivity of SPICA \textit{cooled} telescope.
It~is concluded that only through sensitive far-IR observations much 
beyond \textit{Herschel} capabilities 
we will be able to constrain the mass, the energy budget 
and the water content of hundreds of protostars and planet-forming disks.

\end{abstract}

\section{Introduction}

Modern astrophysics is just beginning to provide answers to some of the most
basic questions about our place in the Universe: Are Solar Systems like our own
common among the millions of stars in the Milky Way and, if so, what implications
does this have for the occurrence of planets that might give rise to life?
The basic building blocks of a planetary system, gas and dust, emit efficiently at 
mid- and far-IR wavelengths ($\sim$5 to 350\,$\mu$m), a critical domain in which to 
unveil the processes that transform 
the interstellar material into stars and planets.

Due to the atmospheric opacity, the far-infrared  domain ($\sim$30-350\,$\mu$m)  has been 
one of the  last spectral windows used in astrophysics. Indeed, observing at these long wavelengths
with sufficient angular resolution and sensitivity represents
a significant technological challenge that requires space telescopes equipped with sophisticated cryogenic instrumentation
(down to a few mK) and very sensitive detector arrays.

The far-IR domain begun to be fully exploited 
by the NED/UK/NASA's IRAS telescope 
and the  ESA's \textit{Infrared Space Observatory}, ISO. 
It has been successfully  continued  by NASA's \textit{Spitzer},  
JAXA's \textit{Akari} 
and most recently by the very successful  \textit{Herschel Space Observatory}, 
the largest telescope ever launched to space and a cornerstone mission of ESA 
with science instruments provided by European-led Principal Investigator consortia and with participation from NASA
\citep[][]{Pilbratt_2010}.

Observations with the above space telescopes revolutionised our understanding of where and how stars are born.
Objects in the early phases of star formation (SF), and
in the first stages of planet formation (protoplanetary disks) release  most of their energy 
in this domain. The far-IR range provides key spectral diagnostics 
physical conditions (atomic fine structure lines, high-$J$ CO lines, etc.), 
chemical evolution (water and light hydrides), 
dust  composition (minerals and ices) and dust column density. 
Therefore, observing in this critical  domain we can 
constrain the mass of protostars and planet-forming disks, access to their most 
important gas cooling lines and ultimately, understand their nature and predict their evolution.

The power of SPICA, a large single-dish \textit{cooled} telescope with unprecedented sensitivity, 
broadband coverage and large field-of-view, 
will drastically improve our understanding  of the star and planet formation processes 
(at small spatial scales), its environment, and  
its link to galaxy evolution (SF at large scales). 
In order to progress in these fields, SED mapping of large  areas of the sky, as well as 
more detailed spectroscopic studies of large samples of young stellar objects (YSOs)
at sensitivities well below those
achieved by \textit{Herschel} are clearly needed. 
In this context, only a very sensitive instrument covering the critical
``far-IR gap'' (\textit{e.g.,} SAFARI) will be able to characterise  the objects
and the environments that are too obscured for the 
the \textit{James Webb Space Telescope} (JWST) to examine in the mid-IR, or too warm
and extended to be efficiently observed by the
the \textit{Atacama Large Millimeter/submillimeter Array} (ALMA).

\section{Lessons from \textit{Herschel} and the need to go beyond}

In this contribution, we  summarise  some of the most  important
achievements of $\textit{Herschel}$ in the field of star and planet formation.
This review is necessarily incomplete and descriptive. The goal is to present
some areas where $\textit{Herschel}$, by observing in the far-IR, has made unique contributions.
We will conclude by discussing its limitations and how only a \textit{cooled} telescope
like SPICA will push the frontiers of our knowledge.

\subsection{The filamentary structure of the ISM}

\textit{Herschel}'s photometric cameras had the ability to map several square degrees in the 
70/100/160\,$\mu$m  (PACS)
and 250/350/500\,$\mu$m (SPIRE) bands. \textit{Herschel} images
confirmed that nearby interstellar clouds are systematically structured in networks of filaments,
the \textit{universal filamentary structure} of the ISM 
\citep[\textit{e.g.,}][]{Andre_2010,Molinari_2010}. 
These filaments are reminiscent of the structures found in large-scale
magneto-hydrodynamic turbulence simulations of the ISM \citep[\textit{e.g.,}][]{Padoan_2001}.
The structure and properties of the diffuse ISM provide the initial conditions for the formation of
dense giant molecular clouds (GMCs) where stars are born. 
The physics that governs the large- and small-scale structure of such diffuse clouds 
is a complex interplay between magnetic fields, turbulence, gravity and thermodynamics.
 
\textit{Herschel} has provided spectacular large-scale images of diffuse clouds before
the onset of SF, 
including high latitude clouds like the Polaris Flare \citep{Miville_2010}. 
They unambiguously reveal the filamentary and clumpy structure of the ISM down to spatial scales 
of $\sim$0.01 pc
(10$''$ angular resolution for a source at $\sim$200\,pc or $\sim$0.1\,pc for a source at  2\,kpc).
Sensitivity and high imaging speed in photometric mode were the key instrumental attributes
of \textit{Herschel} photometric cameras.\\\\

Studying the morphology of a few hundred filaments, most of them regions with on-going SF,
the team of P. Andr\'e et al. concluded that these filaments have the same
characteristic width \citep[$\sim$0.1\,pc;][]{Arzou_2011}.  
Although their origin is not fully understood, this scale
corresponds to the same scale below which turbulence becomes subsonic in diffuse,
non-star forming gas.
These findings clearly connect the fundamental properties and structures of the ISM with the 
regions where stars form.

\subsection{From filaments to cores}

Far-IR photometric surveys of nearby clouds further indicate that cold prestellar cores form primarily along 
dense filaments as they fragment \citep[\textit{e.g.,}][]{Andre_2010,Molinari_2010}.
 From the observational point of view, prestellar cores are detected in lines of sight where the column density of 
material is above a given A$_V$ treshold (roughly above 8 magnitudes of extinction).
With a knowledge of the dust properties and of the dust-to-gas conversion factor, one can translate the observed 
far-IR and submm continuum luminosities into total masses. \textit{Herschel} observations show that despite of their relevance
(prestellar cores collapse gravitationally and form stars) only $\sim$2\,$\%$  of the cloud mass is in these cores.
The ``gas to stars conversion'' is far from being  an efficient  process.
Therefore, most of the cloud mass, $\sim$85\,$\%$, is contained in the extended component at lower extinctions 
($A_V<8$\,mag). These more diffuse regions are dominated by the interstellar turbulence, magnetic
field and UV-irradiation.  These results also mean that 
understanding and characterising observationally the extended component of GMCs
 at large scales has a great relevance.

By studying the mass distribution of the observed prestellar cores, the \textit{Herschel} teams were able 
to plot the so called ``core mass function'' (CMF). The CMF appears to resemble the stellar 
initial mass function (IMF) of stars (meaning of course that stars form from cores).
Yet the CMF is: (1) shifted towards higher masses by a factor of x3 and is not well sampled down to masses 
corresponding to the mean mass of the log-normal IMF ($<$0.3\,$M_{\rm Sun}$) and into the brown dwarf regime 
(\textit{e.g.,} Hennebelle \& Chabrier works). Improved sensitivity in the far-IR domain will be
critical to cover this important stellar population (the most numerous) and to fully understand the
origin of the IMF.

Far-IR observations are extremely important  to understand the first stages of protostellar evolution.
In particular, they are critical to search for the ``holy grail'' of SF, the so-called 
\textit{first hydrostatic cores} (FHSC). Prestellar cores are thought to 
collapse isothermally until become dense and opaque. At this point, the radiation of the central object gets
trapped and the gas is heated up to $\sim$2000\,K. At these temperatures, 
molecular hydrogen (containing the bulk of the core mass) 
starts to be dissociated \citep[see \textit{e.g.,}][]{Com12}.
This is a crucial but short-lived ($\sim$10$^2$-10$^3$\,years) evolutionary stage towards the formation of a Sun-like star.
FHSCs are rare objects and thus they are difficult to identify without mapping very large areas
of several SFRs. 
For nearby SFRs, FHSCs should be visible at 70\,$\mu$m (\textit{e.g.,} with \textit{Herschel}) 
but not at 24\,$\mu$m (\textit{e.g.,} with \textit{Spitzer}).
FHSCs have a peculiar SED  not compatible with a single cold grey-body (like that of a starless/prestellar core)
nor with more evolved SEDs of Class-0 protostars. 
\textit{Herschel} observations of the B1-bs core in Perseus have shown that this is a good FHSC candidate
\citep{Pezzuto_2012}. Broad-band far-IR and submm observations over entire SFRs are clearly needed to
improve the number of detections.

\subsection{From deeply embedded protostars to protoplanetary disks}

Understanding the physical processes involved in the earliest stages of stellar evolution requires 
to study the heating and cooling mechanisms that take place in YSOs.
Beyond photometry, far-IR spectroscopic observations can uniquely probe the evolution of the
physical and chemical structures in YSOs, from warm protostellar envelopes and their outflows to 
planet-forming disks around young stars \citep[\textit{e.g.,}][]{vD_2011}.
 
The evolutionary sequence from Class-0 to Class-II protostars is characterised by 
strong changes in the physical conditions (density, temperature, ionisation...), with the gas 
composition and excitation changing accordingly.
Protostellar evolution is complicated and it is  not  fully understood.
The detection of the most important coolants of the warm/hot gas in YSOs (H$_2$, high-$J$ CO, H$_2$O and O) 
allows a quantitative determination of the energy budget in these objects.
Complete far-IR spectral scans of embedded protostars with \textit{Herschel}
show very rich spectra, with more than a hundred lines detected even at medium spectral resolution
 \citep[][see Fig.~1]{Goico12} 
and with a SED peaking  in the far-IR.
In combination with mid-IR spectra, the entire IR  domain is the key piece 
to characterise the envelope, outflow and disk emission \citep[][]{Herc12}.
The far-IR range is uniquely well suited for the detection of tens of lines from warm water vapour
and to detect water ice features (44/62\,$\mu$m bands).
Following the water abundance from prestellar cores to planet-forming disks is specially relevant, 
both as a unique diagnostic tool of the physical conditions in warm/hot gas and also as the key  molecule 
to define the potential habitability conditions later in planetary systems. 

Detailed models of the far-IR atomic and molecular lines have been used to characterise the main 
heating mechanisms in YSOs: shocks, X-rays, UV radiation, etc. \citep[\textit{e.g.,}][]{vK10}. 
Unfortunately, owing to the large integration times required to cover the complete far-IR range,
 \textit{Herschel}/PACS has only been able to spectroscopically  characterise  a 
small sample of bright protostars \citep[\textit{e.g.,}][]{Karska13,Manoj13,Green13}.
Besides, only for a very few YSOs, these objects have complementary mid-IR spectra (from \textit{Spitzer}).
The above limitations have prevented a more global characterisation of SFRs
 with a much larger statistical meaning.

\subsection{The formation of planetary systems}

Planets form in the accretion disks that develop during the collapse and infall of massive protostellar 
envelopes ($\sim$10,000AU) where stars are born. In this context, 
the formation of planetary systems can be seen as a \textit{by-product} of the SF process. 
However, we still dont fully understand how young gas-rich protoplanetary disks evolve
into planetary systems. Circumstellar disks  are very faint, so far difficult to study spectroscopically.
Hence, we still have  a very incomplete understanding of the formation mechanisms responsible of the great diversity
of extra-solar planetary systems detected so far, and of the peculiar features of our own Solar System.
These, of course, ultimately include the formation of Earths-like rocky planets with significant liquid water 
to support life.   

The basic building blocks of any planet-forming disk (gas and dust) 
radiate predominantly in the far-IR band. Key spectral line diagnostics such as warm water vapour, HD, 
atomic oxygen or dust features
can only be observed in this domain.
For the first time, \textit{Herschel} reached the sensitivity to carry out far-IR spectroscopy in a
few bright protoplanetary disks \citep[\textit{e.g.,} ][]{Thi10,Fed12,Mee12}.
Interesting detections include 
 the discovery of significant
reservoirs of both cold and warm water vapour  in planet-forming disks \citep{Hog11,Riv12}.
Together with the detection of [O {\sc i}]\,63\,$\mu$m (the brightest line in disks), 
much more sensitive detections of water (vapour and ice) in
 hundreds of disks are needed to fully understand the processes that
drive the position of the \textit{snow line}, and thus the
mechanisms that lead to the formation of rocky versus gaseous planets.

Molecular hydrogen is the most abundant gas species in protoplanetary disks (containing
$\sim$90$\%$ of the initial disk mass). However, H$_2$ is a symmetric molecule and it does not  
emit radiation efficiently. Instead, the deuterated isotopologue, HD (with its lowest rotational transitions
at 112, 56 and 37\,$\mu$m -- not observable with JWST or ALMA) turns out to be the most powerful 
tracer of the disk total mass. Depending on the excitation conditions, the far-IR lines of HD can be
 many times more emissive than those of H$_2$. The only \textit{Herschel} detection of
  HD ($J$=1-0 line at 112\,$\mu$m) towards TW Hydrae disk 
is a major discovery and implies a disk mass of more than 0.05 solar masses \citep{Ber13}. 
This is enough to form a planetary system like the Solar System despite the advanced age of TW Hya 
($\sim$10Myr). 
Indeed, HD is also an important constituent of Jupiter-like giant
planets and it is also present in the atmospheres of Uranus and Neptune \citep{Feu13}. 
The detection of HD towards the closest protoplanetary disk
(at $\sim$60\,pc) required 7\,h of integration time and pushed the detection limits of PACS.
A factor $\sim$10 better line sensitivity will be needed to detect HD lines in protoplanetary disks
at the distance of the closest SFRs. This will allow us to carry out a unique survey of disks
with SAFARI and accurately constrain the mass
of a statistically significant sample of planet-forming disks in different SFRs.

\section{SPICA, a new generation far-IR cooled space telescope}

While the very successful \textit{Herschel} mission demonstrates that only in the far-IR  we can answer
fundamental questions on the star and planet formation processes, 
\textit{Herschel} has only shown the tip of the iceberg. This conclusion specially applies to the
spectroscopic characterisation of YSOs and disks. The number of objects studied
was clearly limited. Besides, the \textit{Herschel} telescope was
passively cooled to $\sim$80\,K, thus only offering a modest increase in sensitivity 
compared to previous facilities in the far-IR range.
The lack of sensitivity and of  spectrometers designed to carry out large-mapping also precluded 
the spectroscopic characterisation of ISM clouds and SFRs globally.

In order to better explain the need for a new, much more sensitive telescope, 
one has to take into account the capabilities and scientific merit of the major
astronomical infrastructures of the next decade. In $\sim$2025, ALMA will be in full operations, providing
very high angular and spectral resolution observations in the submm and mm domains. 
ALMA will resolve the inner structures and the outflows of YSOs 
and it will observe planet-forming disks in great detail, 
being more sensitive to the low excitation molecular gas.
ALMA, however, will not have access to the most important cooling lines of the warm/hot gas
(high-$J$ CO, H$_2$O, [C{\sc ii}], [O{\sc i}], ...), it will not observe YSOs at their SED peak 
and it cannot observe mineral and ice features  (the basic ingredients
of a planetary system). Also important, ALMA is inefficient for large-scale mapping.

JWST will provide a major increase in angular resolution and sensitivity in the mid-IR domain,
but JWST only covers the $\lambda$$<$28\,$\mu$m range.
Hence, it does not cover the critical far-IR band. Besides, it does not provide a high spectral resolution
spectrometer in the mid-IR. Finally, JWST is not designed to have a coronagraph with
spectroscopic capabilities  to characterise young exoplanets \textit{directly}.
In the next decade, ELT and TMT-type optical and near-IR telescopes will also be built.
Owing to the extinction at these short wavelengths, the regions where stars form
will remain  obscured.

After \textit{Herschel}'s end of mission, windows of the far-IR domain will be still available
for SOFIA stratospheric observations. However, the angular resolution and the sensitivity
is not better, and key molecules like H$_2$O cannot be observed.
All in all, it is clear that a very sensitive  telescope is 
needed to cover the critical  far-IR wavelength range between JWST and ALMA. In order to go significantly beyond 
\textit{Herschel} capabilities, such a space telescope needs to be specialised both in large-scale spectral 
mapping and in providing detailed spectroscopic characterisation of individual YSOs and disks
(\textit{i.e.,}~observe faint lines, a few 10$^{-19}$\,W\,m$^{-2}$, on top of a $\sim$0.1-10\,Jy continuum).

\subsection{SPICA observations of star and planet forming regions}

SPICA’s large and cold aperture ($\sim$3.2\,m at $<$6\,K) will provide a two order of magnitude 
sensitivity advantage over current far-IR facilities. 
In combination with a new generation of highly sensitive detector arrays, the low telescope 
background will allow us to achieve sky-limited sensitivity over the far-IR range.
In its current design, the SAFARI instrument covers the far-IR band with 
a large, fully-sampled field-of-view of 2$'$$\times$2$'$, both in
photometry (48, 85 and 160\,$\mu$m) and in spectroscopy ($\sim$34-210\,$\mu$m).
The angular resolution of SPICA in the 48\,$\mu$m band is $\sim$4$''$.
As an spectrometer, SAFARI will offer a medium resolution mode ($R$$\sim$2000/4000 at 100/50\,$\mu$m)
and a faster low-resolution mode ($R$$\sim$200/400 at 100/50\,$\mu$m).
The latter one will be enough to efficiently detect the dust SED and the brightest cooling lines in very short
times. Contrary to \textit{Herschel} (very time-consuming and inefficient) this will allow us to map large areas of the sky in
periodic intervals of time, opening new avenues for time-variability studies of  protostars and protoplanetary disks in SFRs.\\\\
\textbf{Transition from diffuse ISM clouds to GMCs}:\\SPICA will  allow us to study the
transition from diffuse clouds to
GMCs where stars are form. By accessing the most important
cooling lines of the ISM ([C {\sc ii}]158\,$\mu$m in particular) and the dust SED, 
we will be able to constrain the gas thermodynamics at large-scales and relate it to the
the formation of filaments and their fragmentation.
Dense filaments in SFRs are embedded in a more diffuse and turbulent medium 
that seems to be driven on larger scales. 
As we have seen, these more quiescent and  extended regions constitute the bulk 
of the GMCs mass and play a critical role in their evolution.
In spite of their relevance, the interfaces between the SF cores and the environment 
are less characterised spectroscopically and thus remain poorly understood.
Many questions still need to be answered.
How are GMCs formed and destroyed? 
How does SF affect their structure, evolution, and lifetime? (\textit{feedback}).
How does the environment determine the SF process? (\textit{fragmentation}, \textit{efficiency} ...)

While complete far-IR spectroscopic surveys of a handful of bright YSOs with \textit{Herschel} 
provided the most complete information on the physical conditions and chemical content
of individual sites of SF (hot cores, protostars, H\,{\sc ii}, etc.), 
they do not place the observations in the context of the large-scale 
emission (their environment). This has great relevance since it is 
the widespread gas and dust that set the initial conditions for SF.  
Contrary to \textit{Herschel}, SPICA/SAFARI will map these faint extended regions 
both in the dust continuum and in  bright cooling lines ([C{\sc ii}], [O{\sc i}], [N{\sc ii}], ...), detecting individual objects
and characterising their environment simultaneously.  
As noted earlier, ground-based submm telescopes such as ALMA cannot access the most important 
cooling lines of the warm/hot gas in YSOs. Nor can they observe the SED peak, which is
essential to determine the dust temperature.\\\\\\
\textbf{Spectral characterisation of protostars and planet-forming disks}:\\
Mapping GMCs across the broad wavelength range of SPICA/SAFARI at 
multi-line mapping speeds orders of magnitude beyond the capabilities of 
\textit{Herschel}, we will be able to completely sample a few appropriate 
SFRs for:\\
\textit{(a)} short-lived  first hydrostatic protostellar cores,\\ 
\textit{(b)} protostars down to 0.01\,$M_{\rm Sun}$ throughout 
the Class 0-I-II stages from 0.1-1\,Myrs,\\
\textit{(c)} planet-forming disks with a broad range of masses and  evolutive stages. 

The simultaneous observation of the complete SED, dust grain features
and of a wealth of atomic fine structure, H$_2$O, high-$J$ CO and HD lines 
will make \textit{SAFARI a unique instrument 
to  constrain the mass, the energy budget and the water content of hundreds of YSOs and disks}. 
Also very important, SPICA/MCS will be able to trace the hot gas and resolve 
the gas kinematics in many YSOs and disks, providing
$R$$\sim$30,000 resolution at the wavelength of key H$_2$ rotational lines and of some mid-IR
vibrational bands from organic molecules. 

In conclusion,
complementing JWST and ALMA observations, SPICA  
will allow us to understand the nature of these objects and 
to critically test star/planet formation and evolution theories.

\subsection{Observing with SPICA/SAFARI}

As an example of the powerful SPICA/SAFARI capabilities \citep{Roe12}, Table 1 shows one
of the few protostars (nearby and bright) for which a complete \textit{Herschel} spectrum exists \citep{Goico12}.
In comparison with \textit{Herschel} (a handful of bright individual objects and a few small maps 
of very bright lines), SAFARI will be able to carry-out unbiased spectral-mapping surveys of entire SFRs.
For the first time in the critical far-IR domain, this will allow us to \textit{characterise hundreds of protostars
and planet-forming disks, together with their environment simultaneously}.

\begin{figure}[!ht]
\begin{center}
   \resizebox{0.85\hsize}{!}{
     \includegraphics*{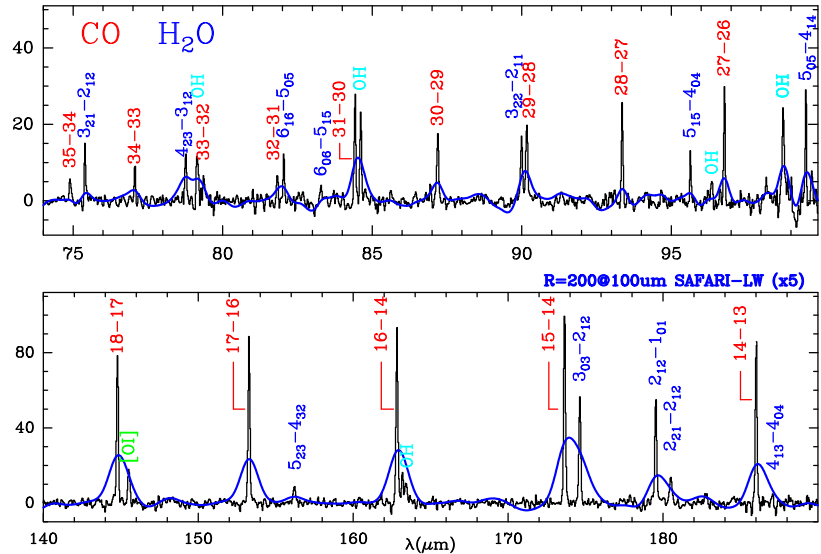}
   }
\end{center}
\caption{Far-IR continuum-subtracted spectrum of the Class~0 protostar Serpens SMM1 taken with
PACS \citep{Goico12}. The blue spectrum shows 
the same spectra as it would be observed in a ``fast'' low-resolution mode with SAFARI
($R$$\sim$200 at 100\,$\mu$m). 
A complete spectrum ($\sim$34-210\,$\mu$m) over a FoV of 2$'$$\times$2$'$ 
could be obtained by SAFARI in only $\sim$30\,sec ($\sim$3\,h and worse sensitivity with PACS).
A higher spectral resolution spectrum (similar to PACS) and better adapted
for the observation of planet-forming disks, can be
obtained by SAFARI in $\sim$4\,min.
}\label{fig:poster}
\end{figure}

\begin{table}[!ht]
\begin{center}
\caption[]{\textit{Herschel} versus SPICA observations of YSOs in the far-IR.\\}\label{tbl:pages}
\begin{tabular}{ccc}
\hline
\hline
Class 0 protostar                         & Serpens SMM1 at 200 pc         & Serpens SMM1 at 2kpc \\
                                          & $\textit{Herschel}$/PACS       & SPICA/SAFARI \\\hline
Continuum at 100\,$\mu$m                  & $\sim$250\,Jy                  & $\sim$2.5\,Jy \\
O{\sc i}\,63\,$\mu$m (brightest line)     & 8E-15 W\,m$^{-2}$              & 8E-17 W\,m$^{-2}$\\
CO $J$=30-29 at 87.2\,$\mu$m              & 7E-16 W\,m$^{-2}$              & 7E-18 W\,m$^{-2}$\\\hline
Complete far-IR spectrum                  & $\sim$3h                       & $\sim$4\,min \\
($R$$\sim$2000 at 100\,$\mu$m)            & (rms$\sim$5E-18\,W\,m$^{-2}$)  & (rms$\sim$1E-18\,W\,m$^{-2}$) \\\hline
10$'$$\times$10$'$ fully sampled map      & $\sim$100\,h for 1 line        & $<$3\,h$^*$ (full far-IR band) \\
($R$$\sim$2000 at 100\,$\mu$m)            &                                & \\\hline
10$'$$\times$10$'$ fully sampled map      & Not possible                   & $<$1.5\,h$^*$ (full far-IR band) \\
($R$$\sim$200 at 100\,$\mu$m)             &                                & \\\hline\hline
\end{tabular}
$^*$Assumes a pessimistic overhead of $\sim$3\,min between each consecutive FoV position.\\
\end{center}
\end{table}



\acknowledgements 
J.R.G. thanks the Spanish MINECO for funding support through grants 
CSD2009-00038, AYA2009-07304, AYA2012-32032 and through a Ram\'on y Cajal research contract. 
We warmly thank the SPICA and SAFARI instrument teams.

\bibliography{goicoechea}

\end{document}